# Weakly superconducting anisotropy in $4H_b$-Nb$_{0.95}$Ti$_{0.05}$Se$_2$ with $1T/1H$ heterostructure

Fanyu Meng[1,2] and Hechang Lei[1,2,*]

[1]*School of Physics and Beijing Key Laboratory of Opto-electronic Functional Materials & Micro-nano Devices, Renmin University of China, Beijing 100872, China*

[2]*Key Laboratory of Quantum State Construction and Manipulation (Ministry of Education), Renmin University of China, Beijing, 100872, China*

## Abstract

Heterostructures of layered transition metal dichalcogenide (TMD) exhibit rich physical properties by the combination of strong electronic correlation effects in $1T$ layer and superconductivity in $1H$ layer. But the limited number of bulk TMD materials with such heterostructures impedes the in-depth understanding of the physical mechanisms behind these properties, as well as research on tuning of these properties. Here, we report a systematic study on physical properties of the $4H_b$-Nb$_{0.95}$Ti$_{0.05}$Se$_2$ single crystals with $1T/1H$ heterostructure. It exhibits a superconducting transition at temperature below 3.3 K. Further analysis indicates that $4H_b$-Nb$_{0.95}$Ti$_{0.05}$Se$_2$ is an intermediately coupled type-II superconductor. Moreover, $4H_b$-Nb$_{0.95}$Ti$_{0.05}$Se$_2$ shows a rather weak superconducting anisotropy (~ 2), distinctly different from other known TMD superconductors with $4H_b$ heterostructure.

[*]Corresponding author: H. C. L. (hlei@ruc.edu.cn)

# Introduction

Layered transition metal dichalcogenides (TMDs) have attracted significant interest in recent years due to the reduced dimensionality originating from weak interlayer van der Waals (vdW) interaction and distinctive properties related to various polymorphs like 1$T$ and 2$H$ phases. They exhibit many fascinating physical properties such as superconductivity (SC), charge density wave (CDW), large magnetoresistance, and nontrivial band topology *etc.* [1–6]. SC and CDW are distinct collective electronic phenomena that intriguingly coexist in 2$H$-phase TMD materials with. The relationship between SC and CDW in these systems is still under debate. For example, the superconducting transition temperature $T_c$ increases from approximately 0.14 K in 2$H$-TaSe$_2$ to 7.2 K in 2$H$-NbSe$_2$ while CDW is weakened in the opposite order, and $T_{CDW}$ decreases from 120 K to 30 K [7–8]. It is generally assumed that their interactions are competitive, but there is also contrary evidence for cooperation in angle resolved spectroscopy (ARPES) studies [9–10]. Additionally, the 1$T$-phase TMDs, especially 1$T$-TaS$_2$, have attracted extensive attention recently because of their series of CDW transitions followed by transitions to various exotic phases, like Mott insulating state, possible quantum spin liquid state, glass-like resonant valence bond state, hidden electron state *etc.* [11–14].

On the other hand, vdW heterostructures stacking different functional layers together via weak interlayer interaction can exhibit some novel phenomena, such as topological SC and anomalous Hall effects, due to the combination of the physical properties of different materials and tunable strength of interlayer coupling [15–16]. It is worth noting that TMD materials with 1$T$/1$H$ heterostructure have become a promising platform for studying such new phenomena. For example, in the few-layer 1$T$/1$H$-TaS$_2$ heterostructure, artificial heavy fermion state can be generated by Kondo coupling between local magnetic moments in the 1$T$ layer and itinerant electrons in 1$H$ layer [17]. In addition, the bulk 4$H_b$-TaS$_2$ with a natural 1$T$/1$H$ heterostructure also exhibits many novel physical properties. For example, muon spin rotation measurements indicate that it is a candidate hosting chiral SC [18]. Scanning tunneling spectroscopy results suggest the existence of a topological nodes in superconducting state [19]. Transport measurements show the in-plane upper critical field $\mu_0H_{c2,ab}$ in bulk $4H_b$-TaS$_2$ single crystals is far above the Pauli limit due to the local inversion symmetry broken in $1H$-TaS$_2$ layers and the relatively weak interlayer interaction [20]. Moreover, further studies suggest that in $4H_b$-TaS$_2$ there may have two-component nematic order, magnetic memory effects and spontaneous vortices [21–23]. Isostructural $4H_b$-TaSe$_2$ also exhibits some exotic behaviors, such as the double-$T_c$-dome under high pressure [24]. Due to these unique properties exhibited in $4H_b$-TaCh$_2$ (Ch = S, Se), the exploration of other TMD systems with such heterostructure becomes a key issue to understand the coupling between different polymorphic structures and provides an opportunity to discover some novel physical phenomena.

In recent, a universal strategy for the synthesis of TMD single crystals with heterostructure through self-assembly has been proposed. Based on this method, a series of TMD bulk materials with heterostructure were synthesized, such as $6R$-NbSe$_{2-x}$Te$_x$, $4H_b$-Nb$_{1-x}$V$_x$Se$_2$ and $4H_b$-Nb$_{1-x}$Ti$_x$Se$_2$ [25]. Motivated by this study, in this work, we have successfully grown $4H_b$-Nb$_{0.95}$Ti$_{0.05}$Se$_2$ single crystals and carry out a systematic study on its physical properties. The experimental results show that $4H_b$-Nb$_{0.95}$Ti$_{0.05}$Se$_2$ exhibits a SC with $T_c \sim 3.3$ K and it is an intermediately coupled BCS superconductor. Moreover, the superconducting anisotropy of $4H_b$-Nb$_{0.95}$Ti$_{0.05}$Se$_2$ single crystal is rather weak, distinctively different from most of TMD superconductors with heterostructure or intercalation.

## Methods

The single crystals of $4H_b$-Nb$_{0.95}$Ti$_{0.05}$Se$_2$ were grown by using the chemical vapor transport method. Stoichiometric amounts of high-purity elements Nb powder (purity 99.95 %), Ti powder (purity 99.9 %), and Se powder (purity 99.99 %) were mixed together and put into a silicon tube with a length of 220 mm and an inner diameter of 16 mm. The tube was pumped down to 0.01 Pa and sealed, and then was placed in a two-zone horizontal tube furnace. The two growth zones were raised slowly to 1103 K and 1073 K and held there for 10 days. Then the quartz tubes were taken out from the furnace and quenched in the water. The typical size of plate-like $4H_b$-$Nb_{0.95}Ti_{0.05}Se_2$ single crystal is approximately $3 \times 2 \times 0.6$ mm$^3$. X-ray diffraction (XRD) patterns were measured using a Bruker D8 X-ray diffractometer with Cu $K_\alpha$ radiation ($\lambda = 0.15418$ nm) at room temperature. Elemental analysis was carried out by using energy-dispersive X-ray spectroscopy (EDX) in an FEI Nano 450 scanning electron microscope. The measurement of Hall resistivity was conducted using a superconducting magnetic system (Cryomagnetics, C-Mag Vari-9) with a standard five-probe configuration. In order to eliminate the influence of voltage probe misalignment, we measured Hall resistivity at positive and negative field up 9 T. The Hall resistivity was obtained by antisymmetrizing raw data. Electrical transports and heat capacity were performed using a Quantum Design physical property measurement system (PPMS-14T). Magnetic properties were performed by the magnetic property measurement system (MPMS3).

## Results and Discussions

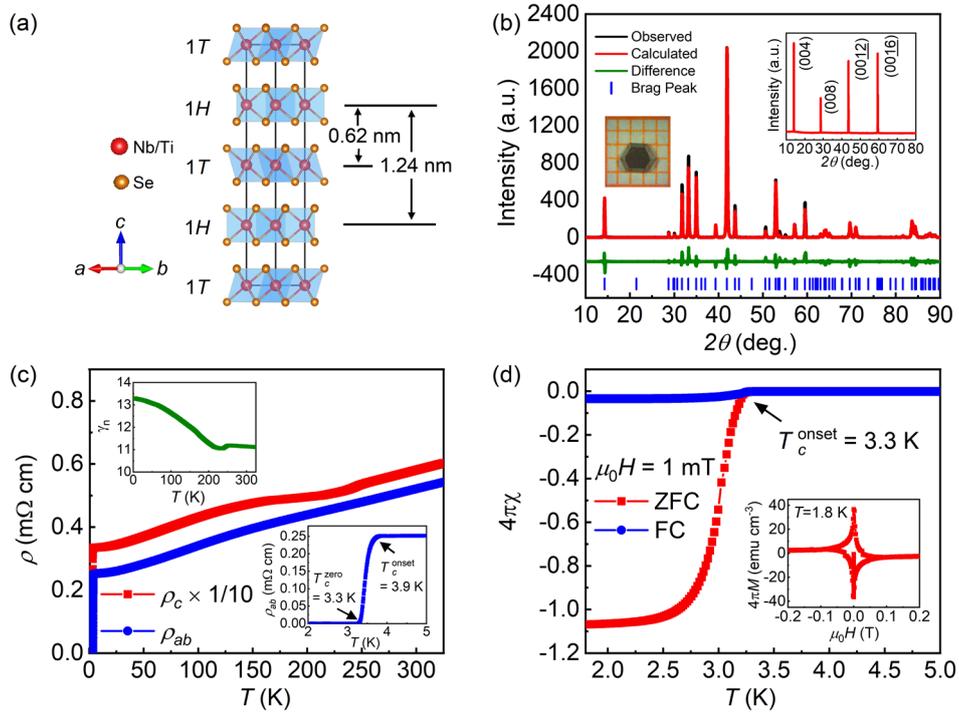

**Figure 1**. (a) Crystal structure of $4H_b$-$Nb_{0.95}Ti_{0.05}Se_2$. (b) Powder XRD pattern of crushed $4H_b$-$Nb_{0.95}Ti_{0.05}Se_2$ single crystals. Insets show the XRD pattern of a $4H_b$-$Nb_{0.95}Ti_{0.05}Se_2$ single crystal (right) and a photo of $4H_b$-$Nb_{0.95}Ti_{0.05}Se_2$ single crystal on a 1 mm grid paper (left). (c) Temperature dependence of zero-field $\rho_{ab}(T)$ and $\rho_c(T)$ for $4H_b$-Nb$_{0.95}$Ti$_{0.05}$Se$_2$ single crystal. Inset: temperature dependence of $\gamma_n = \rho_c(T)/\rho_{ab}(T)$ (upper left) and the enlarged view of $\rho_{ab}(T)$ curve at low-temperature region (lower right). (d) $4\pi\chi$ as a function of $T$ at $\mu_0H = 1$ mT along the $ab$ plane with ZFC and FC modes. Inset shows the magnetization loop at $T = 1.8$ K.

The crystal structure of (Nb, Ti)Se$_2$ can be constructed as a stacked structure of (Nb, Ti)Se$_2$ layers along $c$ axis with weak vdW interactions. The typical structure of NbSe$_2$ is the $2H$-type structure with NbSe$_6$ trigonal prism layer while TiSe$_2$ has the $1T$-type structure with TiSe$_6$ octahedra layer. The $4H_b$-Nb$_{0.95}$Ti$_{0.05}$Se$_2$ is one of (Nb, Ti)Se$_2$ polymorphs, which is composed of alternating stacking of $1H$- and $1T$-(Nb, Ti)Se$_2$ layers [Fig. 1(a)]. The $4H_b$-Nb$_{0.95}$Ti$_{0.05}$Se$_2$ has the hexagonal symmetry with $P6_3/mmc$ space group (No. 194). It is noted that without Ti doping the pure $4H_b$-NbSe$_2$ can not be stabilized at room temperature and $4H_b$-Nb$_{1-x}$Ti$_x$Se$_2$ only have a very narrow phase range with $x = 0.05 – 0.09$ [25–27]. The powder XRD pattern of crushed $4H_b$-Nb$_{0.95}$Ti$_{0.05}$Se$_2$ single crystal can be fitted well by using this structure [Fig. 1(b)]. It confirms the grown crystals are $4H_b$ phase not $1T$ or $2H$ phases. The fitted $a$- and $c$-axial lattice parameters are 0.34601(1) nm and 2.4845(1) nm, which are closed to the reported values in the literature [26, 28]. Correspondingly, the calculated interlayer distance $s$ between two $1H$-(Nb, Ti)Se$_2$ layers is about 1.24 nm, when the interlayer distance between $1H$ and $1T$ layers is about 0.62 nm. The right inset of Fig. 1(b) shows the XRD pattern of a $4H_b$-Nb$_{0.95}$Ti$_{0.05}$Se$_2$ single crystal. All of peaks can be indexed by the indices of (00$l$) planes, confirming that the $c$ axis is perpendicular to the crystal surface. The hexagonal shape of crystal [left inset in Fig. 1(b)] is also consistent with the XRD pattern and the symmetry of the hexagonal crystal. The EDX analysis yields the atomic ratio of Nb : Ti : Se = 0.99(6) : 0.050(2) : 2 when setting the content of Se as 2.

Figure 1(c) shows the temperature dependence of $ab$-plane and $c$-axial resistivity $\rho_{ab}(T)$ and $\rho_c(T)$ for $4H_b$-Nb$_{0.95}$Ti$_{0.05}$Se$_2$ single crystal from 325 K to 2 K at zero field. It can be seen that both $\rho_{ab}(T)$ and $\rho_c(T)$ decrease with decreasing temperature, indicating the metallic behavior of $4H_b$-Nb$_{0.95}$Ti$_{0.05}$Se$_2$. In addition, $\rho_c(T)$ exhibits a distinct inflection point at $T \sim 250$ K when compared to $\rho_{ab}(T)$. The resistivity anisotropy $\gamma_n(T)$ ($= \rho_c(T)/\rho_{ab}(T)$) is $\sim 11.1$ at 325 K. It increases quickly when $T < 250$ K and finally it becomes about 13.3 at $T = 5$ K [upper left inset of Fig. 1(c)]. It is noted that the $\gamma_n(T)$ of 4$H_b$-Nb$_{0.95}$Ti$_{0.05}$Se$_2$ is much smaller than that of 2$H$-NbSe$_2$ ($\sim 206 - 240$) [29]. On the other hand, when lowering temperature further, 4$H_b$-Nb$_{0.95}$Ti$_{0.05}$Se$_2$ exhibits a superconducting transition [lower right inset of Fig. 1(c)]. The onset transition temperature $T_c^{onset}$ is about 3.9 K and the zero-resistivity temperature $T_c^{zero}$ is about 3.3 K with the transition width $\Delta T \sim 0.6$ K. Interestingly, the $T_c^{onset}$ is close to the value of exfoliated monolayer NbSe$_2$ [30], suggesting that the existence of 1$T$ layer in 4$H_b$-Nb$_{0.95}$Ti$_{0.05}$Se$_2$ may decouple the interlayer interaction of 1$H$ layers significantly. The diamagnetic signal in magnetic susceptibility 4$\pi\chi$ curve at $\mu_0H = 1$ mT with zero-field-cooling (ZFC) mode further indicates that the drop of resistivity corresponds to the superconducting transition [Fig. 1(d)]. The estimated $T_c^{onset}$ is about 3.3 K, consistent with the value obtained from $\rho_{ab}(T)$ curve. At $T = 1.8$ K, the superconducting volume fraction estimated from the ZFC 4$\pi\chi$ curve is about 106 %, confirming the bulk nature of SC in 4$H_b$-Nb$_{0.95}$Ti$_{0.05}$Se$_2$. Compared with the ZFC 4$\pi\chi(T)$ curve, a relatively weak diamagnetic signal of field-cooling (FC) 4$\pi\chi(T)$ curve can be ascribed to flux pinning effect, suggesting that 4$H_b$-Nb$_{0.95}$Ti$_{0.05}$Se$_2$ is a type-II superconductor. Moreover, the 4$\pi M(\mu_0H)$ curve [inset of Fig. 1(d)] at 1.8 K shows obvious hysteresis, further confirming the feature of type-II superconductor for 4$H_b$-Nb$_{0.95}$Ti$_{0.05}$Se$_2$.

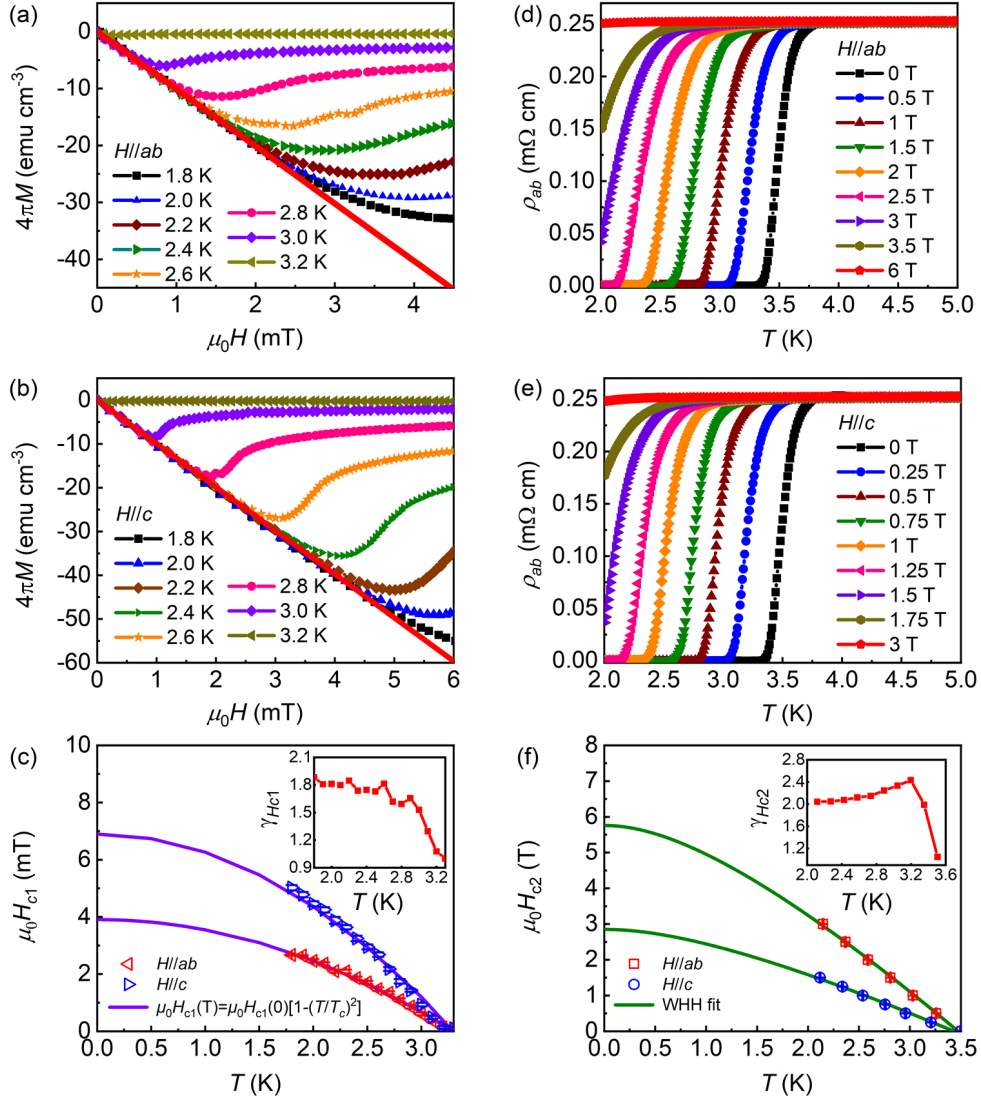

**Figure. 2**. (a) and (b) Low-field parts of $4\pi M$ ($\mu_0 H_{eff}$) curves at various temperatures below $T_c$ for $H//ab$ and $H//c$. The red lines represent the Meissner line. (c) Temperature dependence of derived $\mu_0 H_{c1}(T)$. The purple lines are the fits using the formula $\mu_0 H_{c1}(T) = \mu_0 H_{c1}(0)[1 - (T/T_c)^2]$ for both field directions. Inset: $\gamma_{Hc1}$ as a function temperature. (d) and (e) $\rho_{ab}(T)$ as a function of temperature in various applied magnetic fields for $H//ab$ and $H//c$. (f) Temperature dependence of derived $\mu_0 H_{c2}(T)$. The green lines represent the fits using the WHH formula for both field directions. Inset: temperature dependence of $\gamma_{Hc2}$.

Figures 2(a) and 2(b) show the $4\pi M$ as a function of field at various temperatures below $T_c$ for $H//ab$ and $H//c$. The effective field $\mu_0 H_{eff}$ is calculated by considering the demagnetization effect using the formula $\mu_0 H_{eff} = \mu_0 H_{ext} - N_d M$ [31], where $\mu_0 H_{ext}$ is the external field and $N_d$ is the demagnetization factor. The $4\pi M(\mu_0 H_{eff})$ curves exhibit linear behaviour at low-field region and the slopes are very close to -1, corresponding to the Meissner lines (red lines). Thus, the full Meissner shielding effect in our measurements provides a reliable method for determining the value of $\mu_0H_{c1}$. The $\mu_0H_{c1}$ is determined as the point deviating from linearity based on the criterion $4\pi\Delta M = (4\pi M - 4\pi M_{th}) = 1$ emu cm$^{-3}$, where $4\pi M$ is the measured moment value and $4\pi M_{th}$ ($= -\mu_0H_{eff}$) is the calculated moment value in the same field. The derived values of $\mu_0H_{c1}(T)$ at different temperatures for both field directions are shown in Fig. 2(c). The $\mu_0H_{c1}(T)$ can be fitted well using the formula $\mu_0H_{c1}(T) = \mu_0H_{c1}(0)[1 - (T/T_c)^2]$ (purple lines) and the obtained $\mu_0H_{c1}(0)$ is 3.91(4) mT and 6.9(1) mT for $H//ab$ and $H//c$, respectively. The calculated anisotropy of $\mu_0H_{c1}$, $\gamma_{Hc1} = \mu_0H_{c1,c}/\mu_0H_{c1,ab}$, is shown in the inset of Fig. 2(c). It increases with decreasing temperature in general and the value of $\gamma_{Hc1}$ is ~ 1.88 at $T$ = 1.8 K [inset of Fig. 2(c)], is somewhat smaller than that of 2$H$-NbSe$_2$ (~ 2.97) [32].

The upper critical field $\mu_0H_{c2}$ is one of the most important superconducting parameters, which provides important information about fundamental superconducting properties such as coherence length, anisotropy, the dimension of the superconducting, as well as insights into the mechanism of pair-breaking. The temperature-dependent resistivity $\rho_{ab}(T)$ of 4$H_b$-Nb$_{0.95}$Ti$_{0.05}$Se$_2$ below 5 K in various fields for $H//ab$ and $H//c$ are shown in Figs. 2(d) and 2(e). With increasing field, the $T_c$ shifts to lower temperature gradually in both field directions but the width of superconducting transition only increases slightly, implying the relatively weak flux pinning effect in 4$H_b$-Nb$_{0.95}$Ti$_{0.05}$Se$_2$. The $\mu_0H_{c2}$ is determined using the criterion of 50 % normal state resistivity just above $T_c$ and summarized in Fig. 2(f). The $\mu_0H_{c2}(T)$ curves can be fitted well using the Werthamer-Helfand-Hohenberg (WHH) model (green lines) [33], and the obtained values of $\mu_0H_{c2,ab}(0)$ and $\mu_0H_{c2,c}(0)$ of 4$H_b$-Nb$_{0.95}$Ti$_{0.05}$Se$_2$ is 5.76(8) and 2.85(8) T, respectively. As the Pauli limiting field $\mu_0H_P = 1.86\ T_c$ ~ 6.1 T is larger than the $\mu_0H_{c2,ab}(0)$, it suggests that the orbital-depairing mechanism should be dominant in 4$H_b$-Nb$_{0.95}$Ti$_{0.05}$Se$_2$ [34]. Zero-temperature Ginzberg-Landau (GL) coherence length $\xi_{GL}(0)$ can be estimated with GL formula $\mu_0H_{c2,c}(0) = \Phi_0/[2\pi\xi_{ab}^2(0)]$ and $\mu_0H_{c2,ab}(0) = \Phi_0/[2\pi\xi_{ab}(0)\xi_c(0)]$ [35–37], where $\Phi_0$ is quantum flux ($= h/2e = 2.07\times10^{-15}$ Wb). The calculated $\xi_{GL,ab}(0)$ and $\xi_{GL,c}(0)$ is 10.8(2) nm and 5.3(1) nm, respectively. Based on the values of $\mu_0H_{c1}(0)$ and $\mu_0H_{c2}(0)$, the GL parameter $\kappa_{GL}$ is obtained from the formula $\mu_0H_{c2}(0)/\mu_0H_{c1}(0) = 2\kappa^2/(\ln\kappa + 0.08)$. And the zero-temperature thermodynamic critical field $\mu_0H_c(0)$ can be obtained from $\mu_0H_c(0) = \mu_0H_{c2}(0)/[\sqrt{2}\kappa(0)]$ and the calculated $\mu_0H_c(0)$ is 74(1) mT. The GL penetration length $\lambda_{ab}(0)$ and $\lambda_c(0)$ is 284(8) nm and 605(26) nm, evaluated using the formula $\kappa_c(0) = \lambda_{ab}(0)/\xi_{ab}(0)$, and $\kappa_{ab}(0) = [\lambda_{ab}(0)\lambda_c(0)/\xi_{ab}(0)\xi_c(0)]^{1/2}$ [38]. All of superconducting parameters of $4H_b$-$Nb_{0.95}Ti_{0.05}Se_2$ are listed in Table 1. The anisotropy of $\mu_0H_{c2}$, $\gamma_{Hc2} = \mu_0H_{c2,ab}(T)/\mu_0H_{c2,c}(T)$, is shown in the inset of Fig. 2(f). When decreasing temperature, it increases sharply near $T_c$ and then decreases slowly. The value of $\gamma_{Hc2}$ at $T = 2$ K is $\sim$ 2, which is also smaller than that of $2H$-$NbSe_2$ ($\sim 3.22 - 4.65$) [39–40].

**Table 1.** Superconducting parameters of $4H_b$-$Nb_{0.95}Ti_{0.05}Se_2$.

| $4H_b$-$Nb_{0.95}Ti_{0.05}Se_2$ | $\mu_0H_{c1}(0)$ (mT) | $\mu_0H_{c2}(0)$ (T) | $\mu_0H_c(0)$ (mT) | $\xi(0)$ (nm) | $\lambda(0)$ (nm) | $\kappa(0)$ |
|---|---|---|---|---|---|---|
| ab | 3.91(4) | 5.76(8) | 74(1) | 10.8(2) | 284(8) | 54.8(5) |
| c | 6.9(1) | 2.85(8) | | 5.3(1) | 605(26) | 26.3(5) |

To further investigate the dimensionality of SC in $4H_b$-$Nb_{0.95}Ti_{0.05}Se_2$, we measured the angular dependence of the $\mu_0H_{c2}(\theta)$ at 2 K, where $\theta$ is the angle between magnetic field and the $c$ axis of the crystal. Fig. 3(a) shows the evolution of $\rho_{ab}(\mu_0H)$ as a function of field at different $\theta$ for $4H_b$-$Nb_{0.95}Ti_{0.05}Se_2$ single crystal. When $\theta = 0°$ ($H//c$), SC is suppressed at a relatively low field. With increasing $\theta$, the superconducting transition shifts to higher fields gradually and the $\mu_0H_{c2}(\theta)$ reaches the maximum value at $\theta = 90°$ ($H//ab$), consistent with the larger $\mu_0H_{c2,ab}$ than $\mu_0H_{c2,c}$. Fig. 3(b) shows the angular dependence of $\mu_0H_{c2}(\theta)$ at 2 K extracted from Fig. 3(a) using the criterion of 50 % normal-state resistivity $\rho_{n,ab}(\mu_0H, \theta)$. The feature of $\mu_0H_{c2}(\theta)$ curve is a bell shape near $\theta = 90°$ without a cusp (inset of Fig. 3(b)), suggesting that the SC in $4H_b$-$Nb_{0.95}Ti_{0.05}Se_2$ should be three-dimensional (3D). For 3D superconductors, the angular dependence of $\mu_0H_{c2}(\theta)$ usually can be described by the anisotropic 3D GL model [36–37], $\left(\frac{\mu_0H_{c2}(\theta)\cos(\theta)}{\mu_0H_{c2,c}}\right)^2 + \left(\frac{\mu_0H_{c2}(\theta)\sin(\theta)}{\mu_0H_{c2,ab}}\right)^2 = 1$. The fit using 3D GL model describes the $\mu_0H_{c2}(\theta)$ behavior in general and the fitted $\mu_0H_{c2,c}$ and $\mu_0H_{c2,ab}$ at 2 K are 1.62(2) T and

3.30(2) T. Correspondingly, the calculated $\gamma_{Hc2}$ is 2.04(1). All of these values are in agreement with the values shown in Fig. 2(f). Within the framework of the classical anisotropic GL theory [36, 41], the anisotropic behaviors of layered superconductors can be characterized by their effective mass anisotropy, $\gamma_{Hc2} = \mu_0 H_{c2,ab}/\mu_0 H_{c2,c} = \xi_{ab}/\xi_c = \gamma_m = (m_c^*/m_{ab}^*)^{1/2} = \lambda_c/\lambda_{ab} = \mu_0 H_{c1,c}/\mu_0 H_{c1,ab} = \gamma_{Hc1}$, where $m_c^*$ and $m_{ab}^*$ is the effective mass along the $c$ axis and in the $ab$ plane, respectively. The nearly same values of $\gamma_{Hc1}$ and $\gamma_{Hc2}$ near 2 K (~ 2) valid this relationship and indicated that the anisotropy of effective mass $\gamma_m$ is about 4. On the other hand, according to the Drude model, the $\gamma_n$ at normal state can be expressed as,

$$\gamma_n = \frac{\rho_c}{\rho_{ab}} = \frac{m_c^*}{ne^2\tau_c} / \frac{m_{ab}^*}{ne^2\tau_{ab}} = \frac{\tau_{ab}}{\tau_c} \frac{m_c^*}{m_{ab}^*} \qquad (1)$$

where $n$ is the carrier concentration, $\tau_{ab}$ and $\tau_c$ are the carrier scattering times in the $ab$ plane and along the $c$ axis. Because the $\gamma_n$ at 5 K (~ 13.3) is much larger than those of $\gamma_{Hc1}$ and $\gamma_{Hc2}$, such large value of $\gamma_n$ could be partially ascribed to the anisotropic scattering time $\gamma_\tau = \tau_{ab}/\tau_c \sim 3.3$.

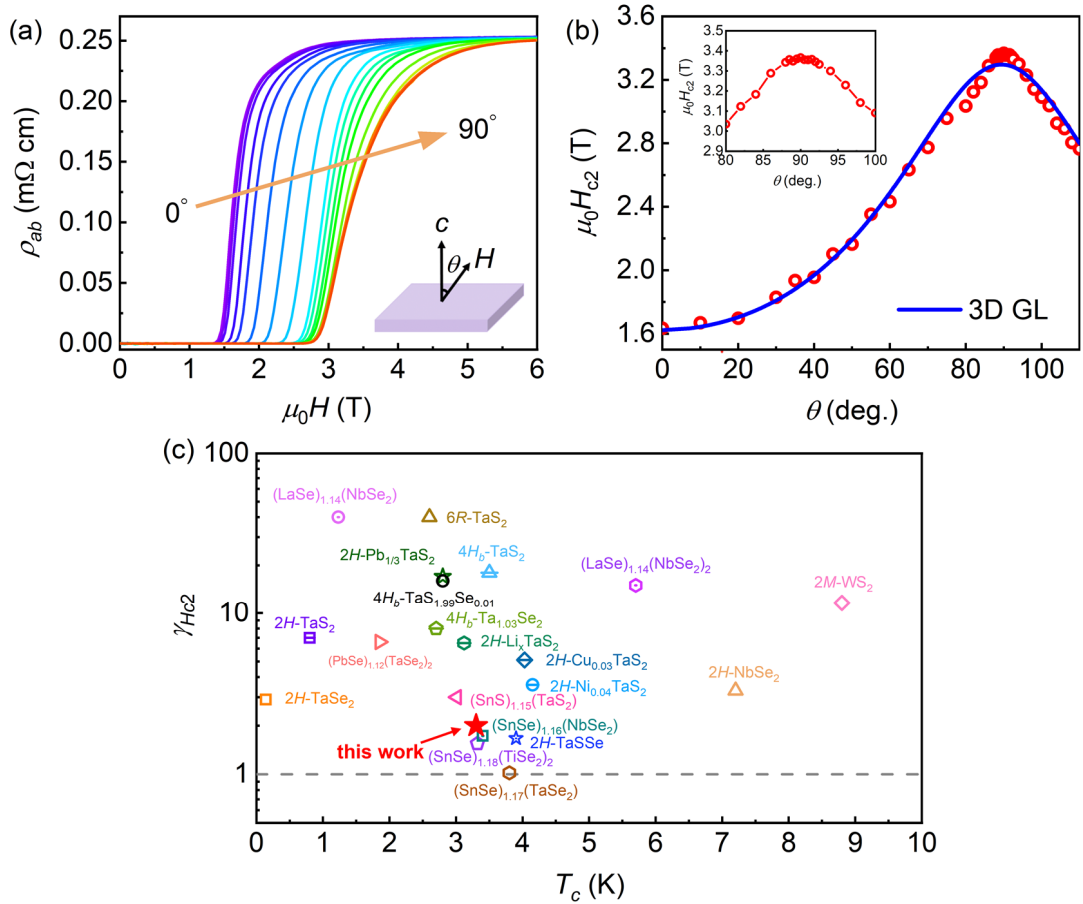

**Figure. 3**. (a) Field dependence of $\rho_{ab}(\mu_0 H)$ at 2 K and various field directions for 4$H_b$-Nb$_{0.95}$Ti$_{0.05}$Se$_2$ single crystal. (b) Angular dependence of $\mu_0 H_{c2}(\theta)$ at 2 K (red circles). The $\mu_0 H_{c2}(\theta)$ at each $\theta$ are determined using the criterion of 50 % $\rho_{n,ab}(\mu_0 H, \theta)$. Inset: enlarged view of $\mu_0 H_{c2}(\theta)$ near $\theta = 90°$ (H//ab). The blue curve represents the fit using 3D anisotropic GL model. (c) The $\gamma_{Hc2}$ comparison of TMD superconductors.

We plot the $\gamma_{Hc2}$ of various TMD superconductors for comparison [Fig. 3(c) and Table S1 in Supplemental Material]. It can be seen that for most of TMD superconductors they have rather large superconducting anisotropy. It is noted that the $\gamma_{Hc2}$ of 4$H_b$-TaCh$_2$ (7.5 – 18) are much larger than that of 4$H_b$-Nb$_{0.95}$Ti$_{0.05}$Se$_2$ [20, 24]. Usually, the larger $\gamma_{Hc2}$ could reflect the evolution of superconducting dimensionality from 3D to two-dimensional (2D) in principle. For 4$H_b$-TaCh$_2$, the relatively weak interlayer interaction and the local inversion symmetry broken in 1$H$-TaCh$_2$ layers lead to the very high $\mu_0 H_{c2,ab}$, far above the $\mu_0 H_P$, and thus the large $\gamma_{Hc2}$ [20]. Above comparison suggests that 4$H_b$-Nb$_{0.95}$Ti$_{0.05}$Se$_2$ may have a stronger interlayer interaction than those in 4$H_b$-TaCh$_2$ and therefore bulk inversion symmetry will weaken Ising SC appearing in monolayer NbSe$_2$ [30] and result in smaller superconducting anisotropy.

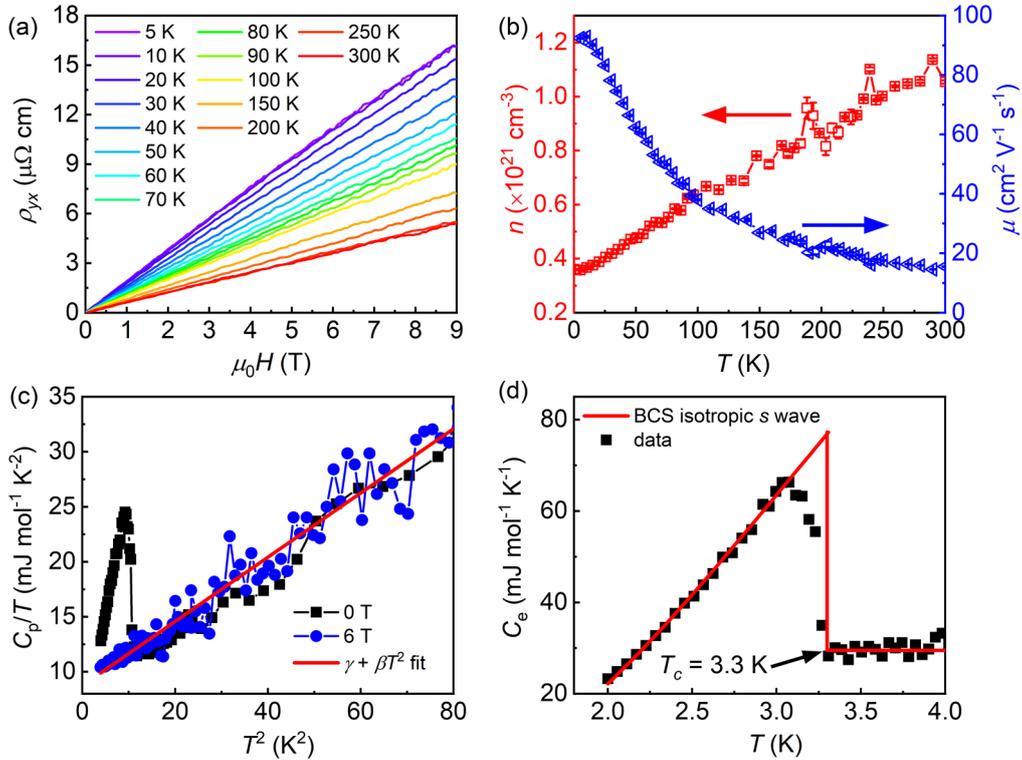

**Figure. 4**. (a) Field dependence of $\rho_{yx}(\mu_0 H)$ at various temperatures. (b) Temperature dependence of derived $n(T)$ and $\mu(T)$. (c) Low-temperature specific heat $C_p/T$ vs. $T^2$ at zero field and 6 T. The red solid line represents the fit of data at 6 T using the formula $C_p/T = \gamma + \beta T^2$. (d) The relationship between $C_e$ and $T$ at zero field. The solid line is the fitted curve assuming an isotropic s-wave BCS gap.

The field dependence of the Hall resistivity $\rho_{yx}(\mu_0 H)$ at various temperatures for $4H_b$-Nb$_{0.95}$Ti$_{0.05}$Se$_2$ single crystal is shown in Fig. 4(a). All curves display a positive slope, suggesting that the hole-type carriers play a dominant role in the transport of $4H_b$-Nb$_{0.95}$Ti$_{0.05}$Se$_2$. Using the linear fits of the $\rho_y(\mu_0 H)$ curves and combining the results of $\rho_{ab}(T)$ at zero field, the carrier concentration $n(T)$ (red squares) and carrier mobility $\mu(T)$ (blue triangles) as functions of temperature can be obtained. As shown in Fig. 4(b), as the temperature decreases from 300 K to 5 K, the $n(T)$ decreases gradually from approximately 1.054(6) to 0.3610(9) × 10$^{21}$ cm$^{-3}$. In contrast, the $\mu(T)$ increases with decreasing temperature from 15.52(9) cm$^2$ V$^{-1}$ s$^{-1}$ at 300 K to 92.4(2) cm$^2$ V$^{-1}$ s$^{-1}$ at 5 K, due to the decrease of electron-phonon scattering.

Figure 4(c) shows the temperature dependence of the specific heat of $4H_b$-Nb$_{0.95}$Ti$_{0.05}$Se$_2$ single crystal below 10 K at $\mu_0 H = 0$ T and 6 T. It can be seen that there is a jump at $T_c \sim 3.3$ K for the curve measured at zero field, confirming the bulk SC of $4H_b$-Nb$_{0.95}$Ti$_{0.05}$Se$_2$. The $T_c$ determined from the specific heat jump is consistent with that obtained from the resistivity and susceptibility. By contrast, at $\mu_0 H = 6$ T, the superconducting transition can not be observed when $T > 2$ K. In order to obtain the electronic specific heat coefficient $\gamma$ and the lattice specific heat coefficients $\beta$, the $C_p/T$ curve at 6 T is fitted using the formula $C_p/T = \gamma + \beta T^2$ (red solid line in Fig. 4(c)). The obtained $\gamma$ is 8.7(3) mJ mol$^{-1}$ K$^{-2}$ when the values of $\beta$ is 0.292(7) mJ mol$^{-1}$ K$^{-4}$. Correspondingly, the calculated Debye temperature $\Theta_D$ is 271(2) K using the equation $\Theta_D = (12\pi^4 NR/5\beta)^{1/3}$, where $N$ (= 3) is the atomic number in the chemical formula and $R$ (= 8.314 J mol$^{-1}$ K$^{-1}$) is the gas constant. The electronic specific heat $C_e$ obtained by subtracting the lattice contribution from the total specific heat is shown in Fig. 4(d). The extracted specific heat jump at $T_c$, $\Delta C_e/\gamma T$ was approximately 1.52, which is a little larger than the value from BCS theory (1.43) [31]. Moreover, the $C_e(T)$ curve below $T_c$ can be fitted using the BCS formula $C_e = a\exp(-\Delta_0/k_B T)$, where $k_B$ is the Boltzmann constant and $\Delta_0$ is the magnitude of the superconducting gap at zero field (red solid line). The fitted value of $\Delta_0$ is 0.544(8) meV and the ratio of $2\Delta_0/k_B T_c$ = 3.83(6), which is also larger than the typical BCS value in the weak-coupling limit (3.53) [42]. On the other hand, the electron-phonon coupling constant $\lambda_{e\text{-}ph}$ can be obtained from the McMillan equation [43],

$$\lambda_{e-ph} = \frac{\mu^* \ln\left(\frac{1.45 T_c}{\Theta_D}\right) - 1.04}{1.04 + \ln\left(\frac{1.45 T_c}{\Theta_D}\right)(1 - 0.62\mu^*)} \qquad (2)$$

where $\mu^* \sim 0.13$ is the common value of the Coulomb pseudo-potential. The calculated $\lambda_{e\text{-}ph}$ is 0.586(2) using $T_c$ = 3.3 K and $\Theta_D$ = 271(2) K. These results indicates that $4H_b$-Nb$_{0.95}$Ti$_{0.05}$Se$_2$ is an intermediately coupled BCS superconductor.

## Conclusion

In summary, we have grown $4H_b$-Nb$_{0.95}$Ti$_{0.05}$Se$_2$ single crystals successfully and studied its physical properties in detail. Experimental results show that $4H_b$-Nb$_{0.95}$Ti$_{0.05}$Se$_2$ is a bulk superconductor with $T_c$ = 3.3 K and has a relatively small superconducting anisotropy (~ 2), possibly due to the large interlayer interaction. Further analysis indicates that $4H_b$-Nb$_{0.95}$Ti$_{0.05}$Se$_2$ is an intermediately coupled type-II BCS superconductor. In contrast to conventional TMD superconductor with heterostructures, $4H_b$-Nb$_{0.95}$Ti$_{0.05}$Se$_2$ with weak anisotropy provides a new platform for investigating the influence of interlayer coupling on properties of both $1T$ and $1H$ layers.

## Acknowledgments

This work was supported by the National Key R&D Program of China (Grants No. 2023YFA1406500 and No. 2022YFA1403800), and the National Natural Science Foundation of China (Grant No. 12274459).